# Modeling Network Architecture: A Cloud Case Study

**Sabah Al-Fedaghi and  Dana Al-Qemlas**

Computer Engineering Department, Kuwait University, Kuwait

**Summary**

The Internet's ability to support a wide range of services depends on the network architecture and theoretical and practical innovations necessary for future networks. Network architecture in this context refers to the structure of a computer network system as well as interactions among its physical components, their configuration, and communication protocols. Various descriptions of architecture have been developed over the years with an unusually large number of superficial icons and symbols. This situation has created a need for more coherent systematic representations of network architecture. This paper is intended to refine the design, analysis, and documentation of network architecture by adopting a conceptual model called a thinging (abstract) machine (TM), which views all components of a network in terms of a single notion: the flow of things in a TM. Since cloud computing has become increasingly popular in the last few years as a model for a shared pool of networks, servers, storage, and applications, we apply the TM to model a real case study of cloud networks. The resultant model introduces an integrated representation of computer networks.

## 1. Introduction

Services such as software as a service, platform as a service and infrastructure as a service are seemingly on a steep growth curve [1]-[2]. The Internet's ability to support a wide range of services depends on the network architecture and its ability to drive innovations necessary for future networks [3]. Network architecture helps build efficient, reliable, cost-effective, and scalable networks to meet present and future requirements [4]. Network architecture commonly refers to abstract principles for the technical design of mechanisms for computer communication [5][6]. As a concept, it is expected to be relatively long-lived and applicable to more than one generation of technology [5]. Its fundamental organization is embodied in its components, their relationships to each other and to the environment, and the principles guiding its design and evolution [7].

In this paper, the focus is on the overall description of network architecture, especially in the context of cloud services. Additionally, we evaluate the efforts to refine the notions of identifying and assembling various elements and specifying the behavior and interactions among the network architecture's structural components. This includes refining design, analysis, and documentation by developing a coherent diagrammatic representation of network architecture. An architectural representation is a collection of artifacts that document the architecture [7]; in this paper, the emphasis is on the diagrammatic method used to express the representation. Current practices require mapping out plans with a network diagram before setting up a network of servers, routers, and firewalls. The process is similar to what a residential architect does (e.g., first understanding the customers' desires for a residence and then incorporating them into the designs and detailed plans) [4]. In this paper, we propose adopting a new modeling methodology called the thinging machine (TM) as a conceptual description of a network's architecture.

A TM is used to model the static description, dynamic behavior, and controls of network systems. The TM model can serve various levels of granularity and complexity by operating as, for instance, a guide for subsequent network specification, analysis, design, and validation. Diagramming the network architecture provides an opportunity to oversee the entire system. Conversely, when something goes wrong, we can use the diagram to troubleshoot matters. Diagrams play a key role in many engineering scenarios involving design analysis, synthesis, collaboration, and education [8]. They have cognitive significance, as they direct work and establish networks of relationships between multiple symbolic fields and are part of an integrative process through which structures appear in the world. Diagrams may be thought of as relays that create meaning and enable symbolic translation from one mode of representation to another [9]. Compared to any simple home architecture diagram, a current network architecture diagram relies on the awkward, arbitrary use of icons and symbols and lacks systemization, as described next.

### 1.1 Problem

Fig. 1 shows some of the 25 collected "interesting network architecture diagrams" from [10]. Fig. 1 reflects many architecture representations that have been developed in recent years with hundreds of superficial icons (see Fig. 2).

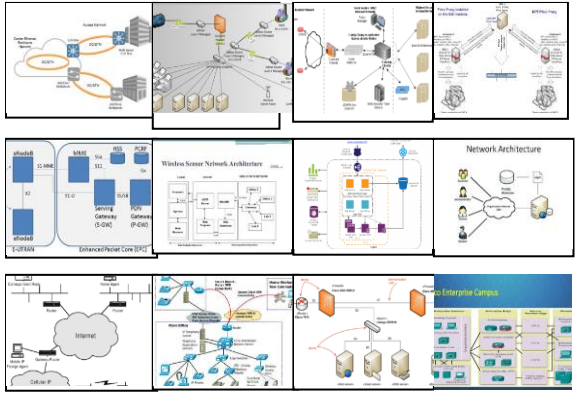

Fig. 1. Samples of network architecture diagrams (From [10]).

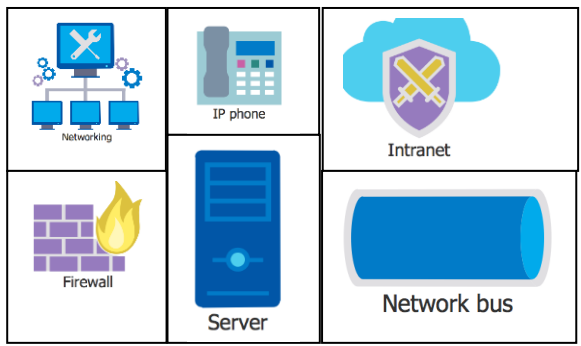

Fig. 2. Samples of proposed network icons (From [11]).

The awkward symbols shown include walls, towers, and human and computer images. Such heterogeneous notions raise the need for more systematic depictions that assist, according to [3], in meeting the challenge of defining a single coherent network architecture.

Even though this paper concentrates on a TM architecture description, TM theory can nevertheless be applied equally to communication diagrams. Network architecture diagrams use unwieldy diagrams with unconstrained use of icons, whereas network communication diagrams go to the other extreme with abstract views of diagrams of nodes and lines. Sometimes, symbols (e.g., a computer screen) are inserted in the representation. The focus in such a description is solely to present the communication occurring between the nodes. This type of model reflects only rough topological connectivity for which mostly unidirectional arrows are used in the network. A meaningful characterization of the individuality of the nodes and a distinction between the static and dynamic aspects of the model are totally absent. Such features will become clear when we use the new TM model in networking. However, with such an extreme level of abstraction, communication diagrams may be needed according to the networking discussed. In this case, the TM can form the foundation of the required rough topological connectivity.

## 1.2 How to Represent a Network

A network is generally viewed in terms of links and a list of who links to whom. This involves how the connection was measured and the kind of link that is utilized. Little attention is paid to the roles of the interiority of a node, which can provide a more fine-grained understanding of networks. In this paper, the TM emphasizes the depiction of the interiority of the nodes' participation in the network while complicating the network. Like a travel map, instead of symbolizing a city with a little circle, the TM represents a node in terms of five generic stages: creating (new things are generated), releasing, transferring, receiving, and processing of the artifacts that flow through and within the nodes. For example, the role of a "dumb terminal" in a network is limited to receiving and releasing artifacts, whereas an intelligent node may create and process things before moving them to other nodes.

The TM views all components of the network in terms of a single notion: movement of things in machines that create, release, transfer, process, and receive. The nodes are specified by their roles, which include creators (i.e., of data), receivers/senders, and processors (reformatters). The TM is applied to a real case study of cloud networks. The general theme of the involved diagramming method is similar to using UML and SysML for modeling systems. However, the UML standard does not have a separate kind of diagram to describe networks, whereas the UML deployment diagrams could be used for this purpose.

To achieve a self-contained paper, Section II reviews the TM that was used in several published papers [12]-[17]. A TM consists of five generic processes of things: creation, processing, releasing, transferring, and receiving, in addition to memory and triggering relations. Section III illustrates TM with a network example that demonstrates the TM's diagraming features by replacing such images as a wall, a cloud, computer servers, etc. with one uniform notation, the TM machine. Section IV applies the TM in an actual case study of cloud networks in a local oil company. The cloud architecture comprises five infrastructures that are (experimentally) remodeled using TM as a conceptual model with operational semantics to produce integration of the static domain description and the dynamic chronology of events. Section IV proposes utilizing TM in simulations; it is suitable for network representation, as it is grounded in a succession of events to reflect the system's behavior.

## 2. Thinging Machines

We adopt a conceptual model [18] centered on things and machines in a network. The philosophical foundation of this approach is based on the German philosopher Heidegger's notion of thinging. Heidegger's philosophy gives an alternative analysis of "(1) eliciting knowledge of routine activities, (2) capturing knowledge from domain experts and (3) representing organizational reality in authentic ways" [19]. Thus, instead of the object-oriented (OO) approach that takes an object as the central concept, we explore alternative process modeling based on the notion of a thing and thinging. In this paper, we describe networks with this process-centered approach that emphasizes operations over objects. It focuses on dynamics, events, and flows rather than static objects.

Thing is a well-defined notion that embraces expansive abstraction instead of the reductive abstraction involved in objects. As will be affirmed in our approach, a thing is also a machine that operates by creating, processing, receiving, releasing, and transferring things. For example, a tree is a thing and a machine through which things (e.g., water, carbon dioxide) flow, and it transforms those flows into various sorts of cells. From this perspective, things are, in their own ways, apparatuses that are not just put to any specific use but are part of a complex mesh of pieces that make up the whole system of interwoven, interacting objects [20]. In contrast, in OO modeling, this picture is wrapped up in a vague connection of class/object, properties, and methods described in terms of multiple diagrams with many kinds of icons and arrows.

The simplest type of thing/machine is referred to by the same name as our model: the TM, a generalization of the known input-process-output model. Things that flow in a TM refer to the exclusive conceptual movement among the five operations (stages) shown in Fig. 3. A thing is what is created, processed, released, transferred, and/or received in a machine. Accordingly, the TM's stages can be described as operations that transform, modify, etc., things in the abstract or concrete sense. They are briefly described as follows.

Arrive: A thing flows to a new machine (e.g., packets arrive at a port in a router).

**Accept**: A thing enters the TM after arrival. We will
**Accept**: A thing enters the TM after arrival. We will assume that all arriving things are accepted; hence, we can combine arrive and accept into the **receiving** stage.
**Release**: A thing is marked as ready to be transferred outside the machine (e.g., in an airport, passengers wait to board after passport clearance).
**Process**: A **thing** is changed.
**Create**: A new thing is born in a machine (e.g., a forward packet is generated in a machine). The term *create* comes from creativity with respect to a system; for instance, constructed things come from already created things or emergent things appear from somewhere.
**Transfer**: A thing is inputted or outputted to/from' a machine.

The TM includes one additional notation—triggering (denoted by a dashed arrow)—which initiates a new flow (e.g., a flow of electricity triggers a flow of air). Multiple machines can interact with each other through flows or by triggering stages.

## 3. Examples of the TM Modeling

In this section, we give examples that demonstrate the TM's diagraming features by rediagraming network diagrams. Note that a TM (Fig. 3) is the tool by which to construct larger machines.

### 3.1 Network Architecture

Fig. 4 includes such images as a wall, a cloud, computer servers, an instrument, and a cylinder with four arrows. Fig. 5 shows the corresponding TM representation in which all elements of the network are represented as machines.

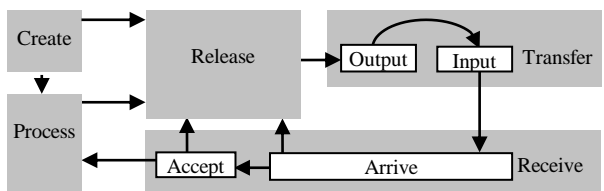

Fig. 3. Thinging machine.

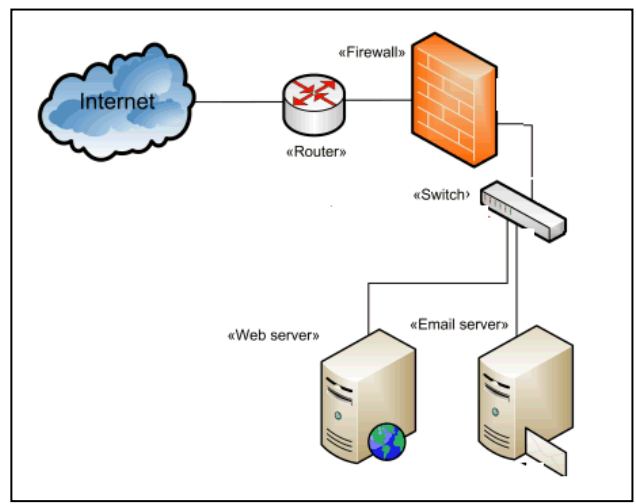

Fig. 4. Network architecture diagram (adapted partially from [21])

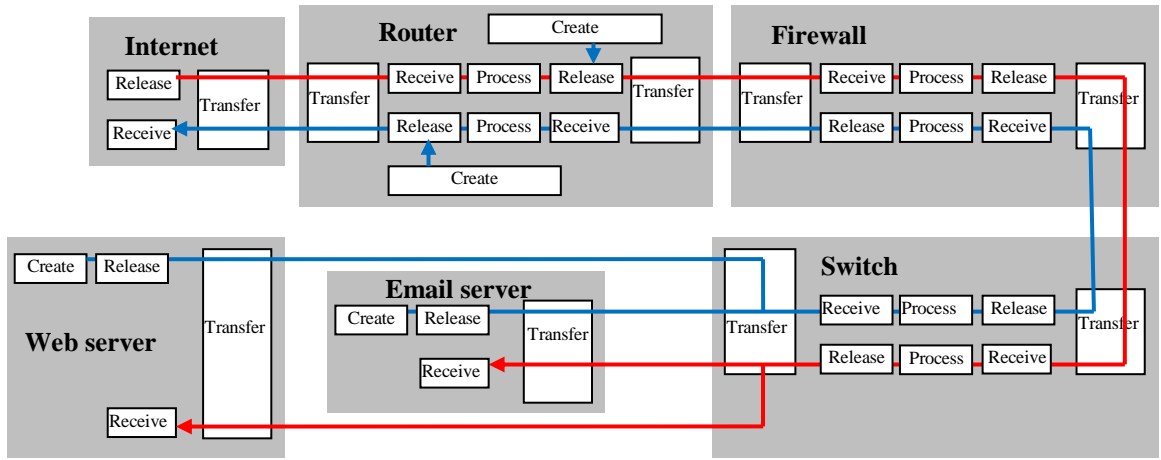

Fig. 5. TM representation of the network architecture diagram in Fig. 4

The arrows represent the streams of packet flows from the servers to the Internet and vice versa; note that Create denotes the construction of new header content in the packet.

First, we demonstrate that the TM is susceptible to various levels of description granularity. Fig. 5 can be simplified by assuming that the direction of the arrow can eliminate the need for the release, transfer, and receive stages. Eliminating the create and process stages can help achieve further simplification. At the end, producing a common method to represent network architecture (as in Fig. 5) is possible by combining the input and output flows of packets and changing the boxes to icons. However, the opposite direction is also possible: here, the TM representation gives more details inside each component in the network, such as the level of a LAN or Ethernet connection. A TM representation can be applied to all internal descriptions of all network components, such as gateways, routers, switches, bridges, and hops. Even the software (e.g., routing algorithms) inside the hardware components of the network can be specified using a TM.

## 3.2 Router

Routing refers to the process of finding a path between two nodes in a network based on certain criteria. Assuming that packet switching occurs, the packets travel from the source to the destination. The router receives the packet and looks at the destination IP address in the packet. It determines the best match for the destination IP address in its routing table. Then, it forwards the packet to that network. For simplicity's sake, we ignore here such details as checksum calculation, maintenance of the routing table, router settings, and queueing of the packets. Additionally, we modeled the router flow in one direction because the other direction can be modeled similarly.

Fig. 6 shows the basic router TM model. In the figure, the packet is received (circle 1) and processed (2) to extract the destination IP (3) and the source IP (4). These two IPs, in addition to data from the routing table (5-8), are processed to produce the next IP (9). Before forwarding and detailing other components of a given network, it is beneficial to look at some of the features of TM modeling as applied to the router represented in Fig. 6. Fig. 6 reflects a static potential machine that can be actualized though dynamic behavior.

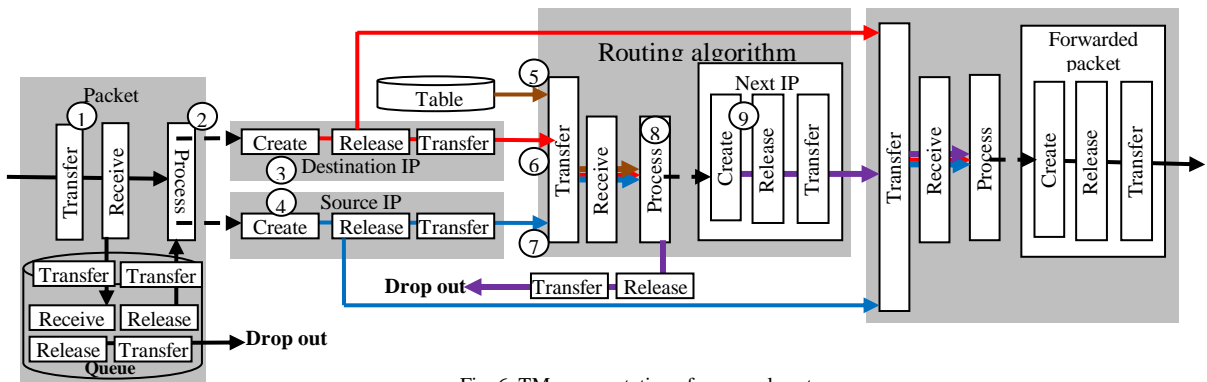

Fig. 6. TM representation of a general router

In a TM, the behavior of the router is captured through events. An event is a TM that comprises time, a region (space field), and the event itself. It may include other properties of the event (e.g., intensity) and implicitly includes the thing that flows in the region. It is a type of a higher-order representation because it includes "chunks" of the stationary representation.

For example, Fig. 7 shows the TM representation of the event of the arrival of a packet. The processing of an event refers to the event takes course. For the sake of simplicity, events will be represented only by their regions. We will assume that after releasing the packet, the next packet on the top of the queue is fetched to be processed. Accordingly, we can designate the following as meaningful events in the router (see Fig. 8):

- Event1 (E1): A packet is received.
- Event2 (E2): The packet is processed.
- Event3 (E3): The packet is queued.
- Event4 (E4): The packet is taken out of the queue.
- Event5 (E5): The packet is dropped.
- Event6 (E6): The source and destination IPs are extracted.
- Event7 (E7): The source and destination IPs and the table are sent to the routing algorithm.
- Event8 (E8): The source and destination IPs are processed to the table.
- Event9 (E9): The forwarding IP is generated.
- Event10 (E10): The source, destination, and forwarded IPs are inputted into the procedure that reconstructed the packet.
- Event11 (E11): The packet is constructed and forwarded to the next node.

Fig. 9 shows the chronology of these events. Note that control can be applied as yet a further type of higher-order representation to include relationships among the events of Fig. 9. As will be discussed in the last section, this model of behavior can be used to simulate the system.

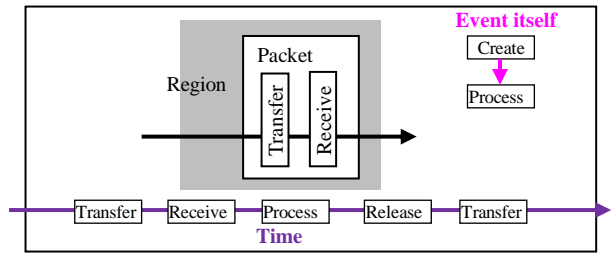

Fig. 7. Event: *Arrival of a packet*

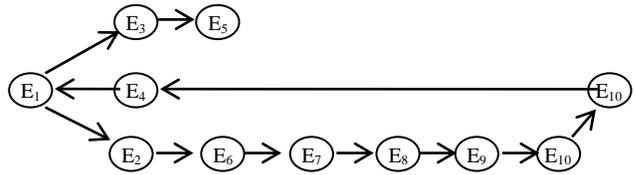

Fig. 9. Events in the TM representation of a general router

## 3.3 Firewall

Without loss of generality, we assume a simple packet-filtering firewall. Fig. 10 shows its TM representation. The firewall is initially in the waiting state until it establishes a connection and the packets start to arrive (3) and be processed (4). Accordingly, the header of the processed packet is extracted (5). Then, the header and the security policy are processed (6, 7, and 8) to generate a decision (9).

- If the decision is positive (10), then the packet is permitted to flow to (11).
- If the decision is negative (12), then the packet is dropped (13), the connection is terminated (14), and the firewall enters a waiting state (15).

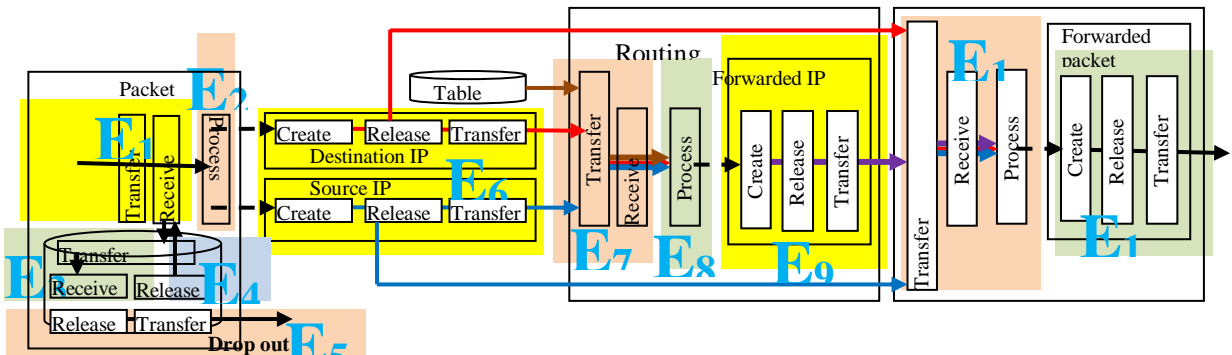

Fig. 8. Events in the TM representation of a general router

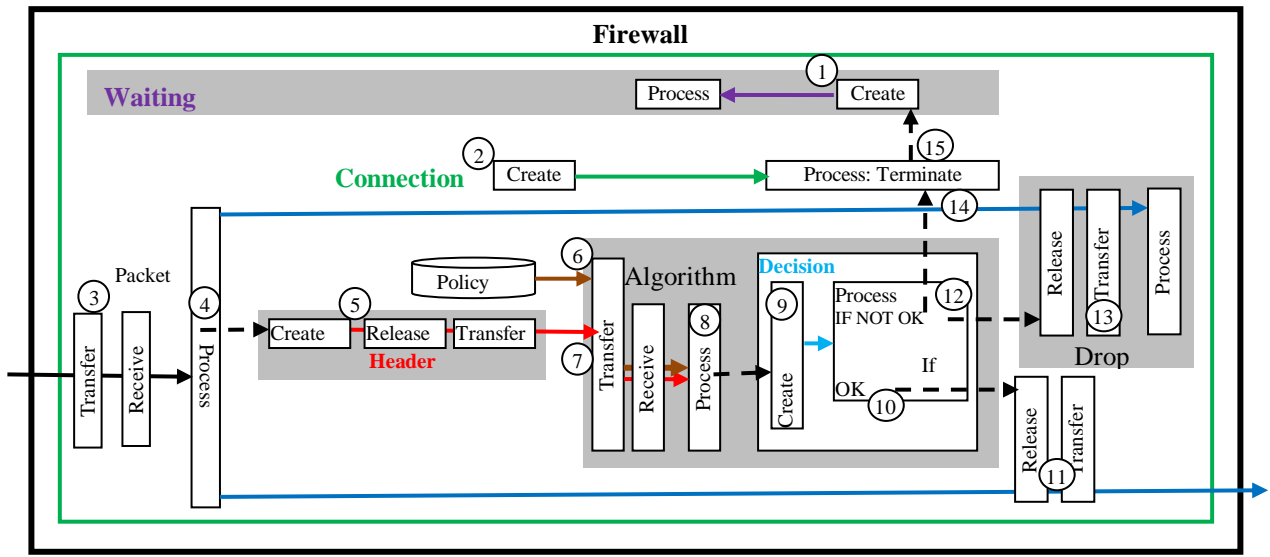

Fig. 10. TM representation of a general firewall

Accordingly, we can designate, as in the case of a router, the following as meaningful events in the physical firewall.

- Event 1 (E1): The firewall is in the waiting state.
- Event 2 (E2): A connection is established.
- Event 3 (E3): A packet arrives and is processed.
- Event 4 (E4): The header is extracted.
- Event 5 (E5): The header is processed according to the policy.
- Event 6 (E6): The result is positive.
- Event 7 (E7): The packet is permitted to proceed.
- Event 8 (E8): The result is negative.
- Event 9 (E9): The packet is dropped.
- Event 10 (E10): The connection is terminated.
- Event 11 (E11): A waiting state is created.
- The corresponding TM for firewall events is not shown due to space limitations.

## 4. Cloud Computing Case Study

Cloud computing has become increasingly popular in the last few years as more technologies are moving to the cloud. "Cloud computing is a model for enabling ubiquitous, convenient, on-demand network access to a shared pool of configurable computing resources (e.g., networks, servers, storage, applications, and services) that can be rapidly provisioned and released with minimal management effort or service provider interaction" [22]. Migrating to the cloud is the future trend for businesses, because to succeed in a competitive business environment, companies require efficient and effective information technology solutions at a low cost [23].

Enterprises are able to attain efficient use of IT hardware and software investments through cloud computing by managing a group of systems as one unit. Implementing cloud computing in enterprises has several advantages and disadvantages with respect to cost, data availability, and data security.

Our case study is a private cloud [24] provisioned and used by a single company (the workplace of the second author). The company's business requirements are growing, causing an increase in the resources of the information technology (IT) department. The IT department requires a large number of servers and storage capacity to accommodate the new demands that require a large amount of space and budget. The cloud computing solution will help accommodate these new requirements while minimizing costs and security concerns [23]. The solution is to apply cloud computing technology to the company. The proposed private cloud would provide centralized, collaborative, efficient, and secure services to the entire company.

The main goal of building a cloud is to construct a network that can cope with demands, enhance service, and lay the groundwork for fast execution of a company's operations team plans to meet the expectations of its user base. In addition, it provides clear financial and high operational visibility to higher management. Cloud implementation requirements include fast service execution through automation, service-oriented IT, showing back charges and chargebacks to users, logging, and monitoring, as well as enhanced security.

## 4.1 Design

The cloud project was designed using the NIST cloud computing model [25], a tool providing the requirements, structures, and processes of cloud computing. The model is composed of five essential characteristics, three service models, and four deployment models [25]. The cloud project includes five essential NIST cloud characteristics. The characteristics include on-demand self-service, because an end user can independently acquire computing capabilities; second, broad network access is included, as capabilities are accessible over the network. Another characteristic of the cloud is resource pooling, given that a multitenant model is used to pool computing resources to multiple clients. In addition, rapid elasticity as a capability can be scaled in or out depending on the requirements. Lastly, the cloud includes measured services to provide transparency by monitoring, controlling, and reporting the resource usage [25].

The cloud includes the following six implemented main service models:

1. An IaaS that provides the network, storage, computations, and operating system components as a service. The IaaS accommodates Windows servers and Red Hat Enterprise Linux servers.
2. Storage as a service that provides the new data stores for virtualization platforms and attaches the storage to the physical infrastructure.
3. Backup as a service that provides different backup and restore functions for the physical machines, virtual machines, and SQL databases.
4. PaaS that includes the SQL server, SQL cluster, and database as a service.
5. Network as a service that covers the enterprise-wide area network and includes load balancer as a service.
6. Disaster recovery as a service that protects applications and data from disruption by using cloud resources to provide an organization with business continuity in the event of a system failure.

The selection of a cloud computing model is a major decision, as each type of cloud has its advantages and disadvantages. The selected cloud deployment model in our case study is a private cloud; it offers increased security and privacy. In addition, the cloud is customizable to meet business and technical needs. Furthermore, the private cloud saves costs compared to public clouds in the long run because its total cost of ownership is lower. Moreover, because the cloud is hosted in the company's private environment, managers have more control over the data. Lastly, the private cloud is owned by the enterprise, ensuring business continuity.

## 4.2 Architecture

The proposed cloud architecture comprises five infrastructures: unified data exchange, information, supervision, networking, and security management platforms. The architecture allows effective communication and resource sharing with other companies. To achieve successful cloud infrastructure planning, designing and standardization are required. The cloud architecture shown in Fig. 11 implements a layered deployment mode. The first layer is the company's departments, and the bottom level is the company's data centers. The scheduling and resource allocation are centralized to enhance resource utilization and increase the speed of service delivery [26].

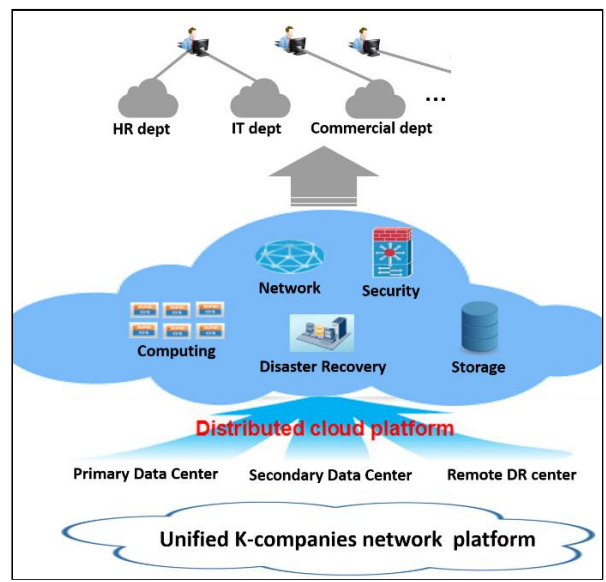

Fig. 11. Cloud architecture (partially redrawn from [26])

## 4.3 TM Modeling

Fig. 12 shows the TM representation of the cloud computing system discussed in the previous section. The user (1) opens the cloud server portal (2) and selects one of the following three options: Server request, Database request and Storage request.

**Server Request**

If the user selects the server request (3), the server request window opens and the user enters the required details: type (production or development), site (main, refinery), environment (Windows or Linux), deployment size (small, medium, large, or custom), application tier (app, web, or DB), and backup (regular or critical).

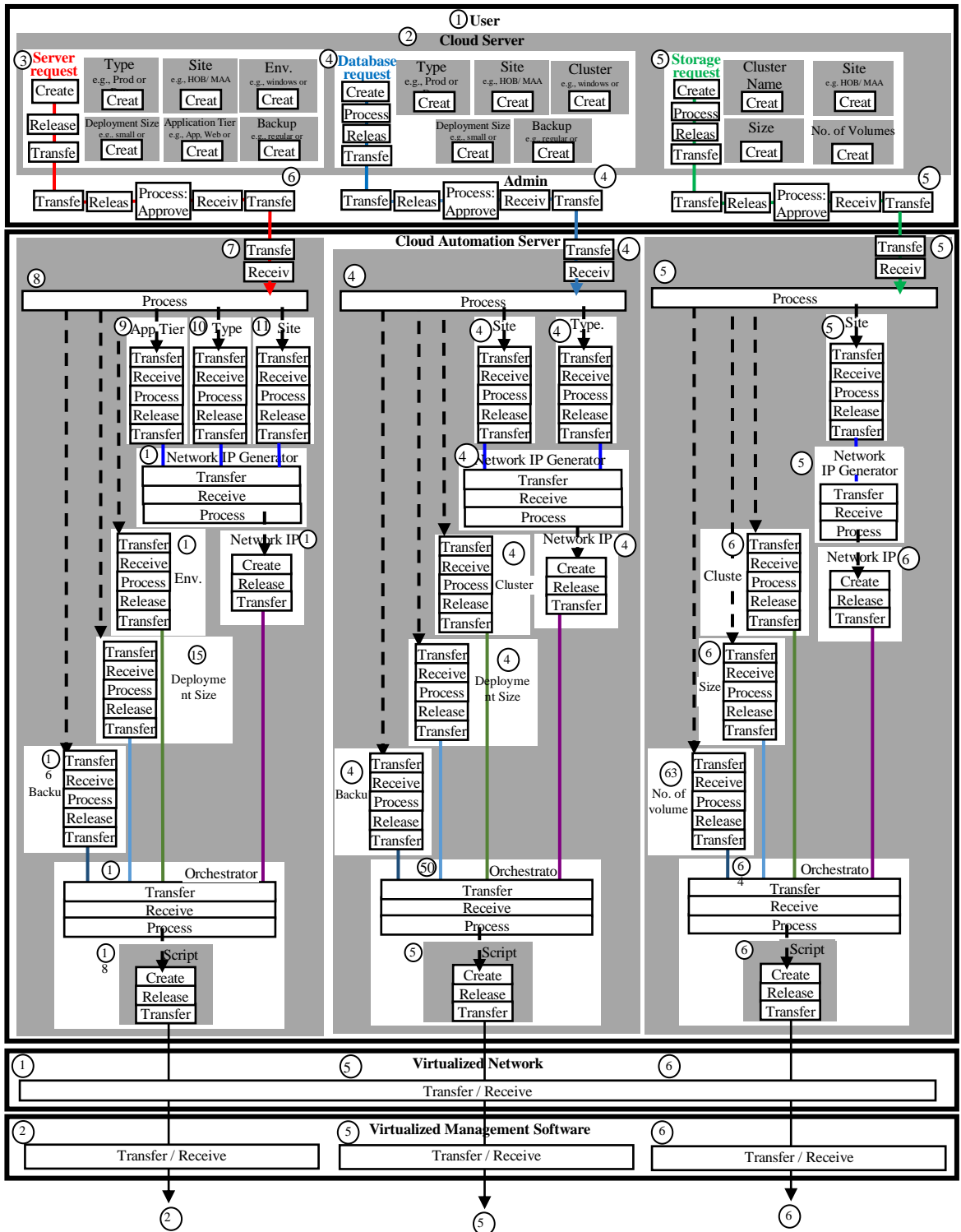

Fig. 12. Cloud TM model

The server request is then transferred to the administrator (6), and the administrator approves the request. The server request is then transferred to the cloud automation tool (7), which processes the details received (8). The app tier (9), type (10), and site (11) are extracted to the network IP generator (12) to trigger the creation of the network IP (13) and are then sent to the orchestrator. The server request is then transferred to the cloud automation tool (7). The cloud automation tool processes the details received (8). The app tier (9), type (10), and site (11) are extracted to the network IP generator (12) to trigger the creation of the network IP (13) and are then sent to the orchestrator.

The server environment (14), deployment size (15), and backup (16) are released and processed by the cloud automation tool and sent to the orchestrator. The orchestrator (17) receives the data, creates a script with the server specifications (18), and sends it to the virtual network. The virtual network receives that script and transfers it to the virtual management software (19). The virtual management software (20) receives the data and processes it (21), as shown in Fig. 13.

### Database Request

If the user selects a database request (4), the database request window opens and the user enters the required details: cluster name, site (main or refined), environment (Windows or Linux), deployment size (small, medium, large, or custom), and backup priority (regular or critical).

The database request is then transferred to the administrator (40), and the administrator approves the request. The server request is then transferred to the cloud automation tool (41). The cloud automation tool processes the details received (42). The site (43) and type (44) are extracted to the network IP generator (45) to trigger the creation of the network IP (46) and are then sent to the virtual network. The cluster (47), deployment size (48), and backup (49) are released and processed by the cloud automation tool and sent to the orchestrator.

The orchestrator (50) receives the data, creates a script with the database specifications (51), and sends it to the virtual network. The virtual network receives that script and transfers it to the virtual management software (52). The virtual management software (53) receives and processes the data (54).

### Storage Request

If the user selects a storage request (5), the server request window opens and the user enters the required details: cluster name, site (main or refinery), environment (Windows or Linux), size (small, medium, large, or custom), and number of volumes. The server request is then transferred to the administrator (55) and the administrator approves the request. The storage request is then transferred to the cloud automation tool (56). The cloud automation tool processes the details received (57). The site (58) is extracted to the network IP generator (59) to trigger the creation of the network IP (60) and is then sent to the virtual network.

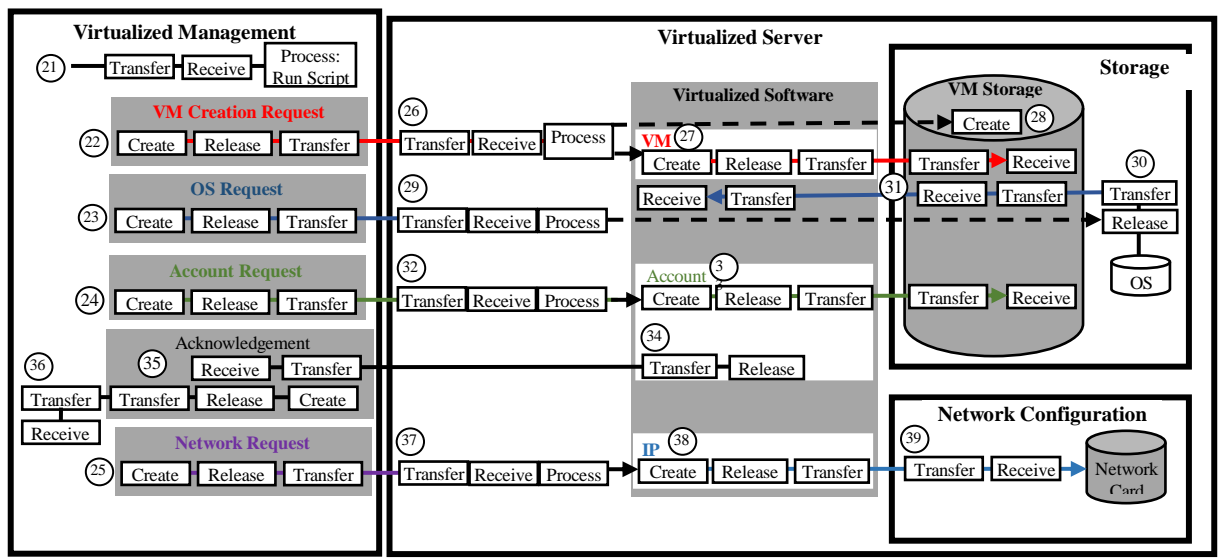

Fig. 13. Server request TM model

The cluster (61), size (62), and number of volumes (63) are released and processed by the cloud automation tool and are then sent to the orchestrator. The orchestrator (64) receives the data, creates a script with the storage specifications (65), and sends it to the virtual network. The virtual network receives that script and transfers it to the virtual management software (66). The virtual management software (67) receives and processes the data (68).

Fig. 13 is the continuation of Fig. 12 and shows the TM representation of the process of creating a virtual server. In the figure, the virtualized management software receives the script and then runs the script (21). The script will then trigger the creation of a VM (virtual machine - 22), OS (operating system - 23), account (24), and network (25).

The VM creation request (22) is transferred to the virtualized server (26), and a VM (27) and the VM storage (28) are created on the server. The OS request (23) is transferred to the virtualized software (29), which triggers the OS to be downloaded from the storage (30) to the VM storage, and then the OS is installed on the server. The account request (24) is transferred to the virtualized software (32), which triggers the creation of a username and password (33) that are stored in the VM storage. The account creation triggers the creation of an acknowledgment (34) that is transferred to the virtualized management software (35) and then to the user (36). The network request (25) is transferred to the virtualized software (37), which triggers the creation of an IP (38). This is then configured in the network card (39).

The resultant conceptual machine as shown in Fig. 12 can be used as a foundation for everything related to the documentation, management, and design of a private cloud network. A good understanding of the problem and the system under examination begins with detailed model formulations [27]:

> Model formulation does not mean a computer program. You should instead use conceptual modeling tools: conceptual diagrams, flow charts, etc. prior to any use of software to implement a model. The purpose of conceptual modeling tools is to convey a more detailed system description so that the model may be translated into a computer representation. General descriptions help to highlight the areas and processes of the system that the model will simulate. [27]

Some relevant diagramming constructs include context diagrams, activity diagrams, and software engineering diagrams that capture the basic aspects and behaviors of a system [27]. The TM can replace all these representations in static and dynamic forms.

In the next section, we give an example of utilizing a TM to simulate a system.

## 4.4 Simulation

Network simulation includes simulation models and methodologies such as discrete event simulation. This modeling technique is suitable for network representation, as it is grounded in a succession of events over time to reflect the system's behavior [28], [29]. Graphical languages such as UML and SysML are typically used to build systems models. These modeling languages include heterogeneous lexicons (class diagram, activity diagram, bloc diagram, sequence diagram, etc.). There is no uniformity and no clear mapping between static and dynamic features of the system. A diagram such as an activity diagram (a type of flowchart) can be viewed as describing the behavior of the system. Over years of simulation research, many diagrammatic methodologies have been utilized in building models in simulation (e.g., activity cycle diagrams [state diagrams], event graphs, Petri nets, control flow graphs [30], UML, and BPMN). Simulation is often based on some type of model of the evolved portion of the world being studied. The underlying model is a static description, typically a type of flowchart (e.g., Arena). The simulation itself is executed by generating events or dynamic aspects into the flowchart specification. Therefore, direct simulation and diagrammatic modeling languages of a system use flowcharting as a basis to describe the behavior of the system. The problem with such an approach is that flowcharts have been the target of complaints regarding their value in design and education [31]. This has led to their near-elimination in computer science. Currently, they have been revived in the form of UML activity diagrams.

The behavior of any system in simulation or modeling depends on the notion of event. Clearly, flowcharting does not incorporate this notion, so users of flowcharts have to find some way to incorporate events in flowcharts. In the simulation language Arena, for example, users declare special events with language such as create, wait, and process and develop a type of flowchart that is suitable for simulation. We can see here a conceptual mix among notions such as processes and events. An event is defined in a TM as a change in a region of the system in the flow of time. For example, "wait" in a TM is the process of time. We propose here that a TM provides a more systematic base for simulation.

A TM can be used as a conceptual model in simulation; specifically, TM operational semantics using events define fine-grained activities resulting in an integration of the static domain description and the dynamic chronology of events. We claim that the TM methodology provides a more complete specification suitable for a static domain model and its dynamic aspects needed for simulation. Without loss of generality, we will discuss flowcharting in the simulation language Arena to focus the incorporation of the TM into simulation [27]. In Arena, the conceptual modeling process includes flowcharting, in which the emphasis is placed on developing a specification for a model that could be implemented in any simulation language environment. The TM diagram can play a key role in the design of a discrete-event simulation model. The TM diagram is transformed into a model of the Arena simulation environment, then simulation parameters are tuned to run simulation experiments. Fig. 14 shows an Arena flowchart of the server request discussed previously in section IV.3. A cognitive agent, which generates this Arena flowchart, depends on how well the designer divides the events of the system and projects them in terms of the Arena flowchart.

For example, a graduate student who did not know the TM model was asked to produce Fig. 14 as an Arena flowchart of the server request system. The flowchart was captured by pure arbitrary use of different sizes of events because events are subjective. Note that the student viewed the event process as a type of verb, or as a series of activities. According to [32], a process is "a set of interrelated or interacting activities which transforms inputs into outputs." Verbs are used to describe steps in a process (activities), and nouns are used to describe items outputted by activities to become input for other activities. In contrast, the TM diagram is an objective capture of all elementary events of the modeled system based on the TM's five verbs.

For example, the mere flow of a thing in a TM consists of three elementary events as shown in Fig. 15: transferring (e.g., arrival to input port), receiving, and storing. So, if we desire a certain level of a coarse description of events (e.g., those sent to replace the TM: release, transfer and receive), we can start with the static TM diagram and identify boundaries of large events from elementary events until we reach the required level of granularity. Consequently, the TM description of events can produce all diagrams where elementary events form larger events. For example, the Arena flowchart in Fig. 14 can be produced from the TM as follows:

    a.  Start with the TM diagram.
    b.  Consider every change (stage) as an event.
    c.  Merge events according to the required level of abstraction.

Accordingly, as shown in Fig. 16, we can designate the events in a request to create a server using the cloud automation tool. Fig. 17 shows the chronology of these events.

- Event 1(E1): The manager receives a server request.
- Event 2 (E2): The manager approves the request.
- Event 3 (E3): The cloud automation tool receives the server request.

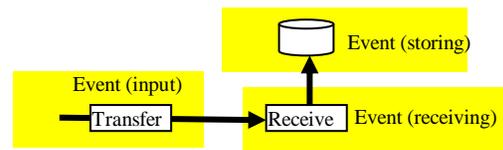

Fig. 15. Elementary events

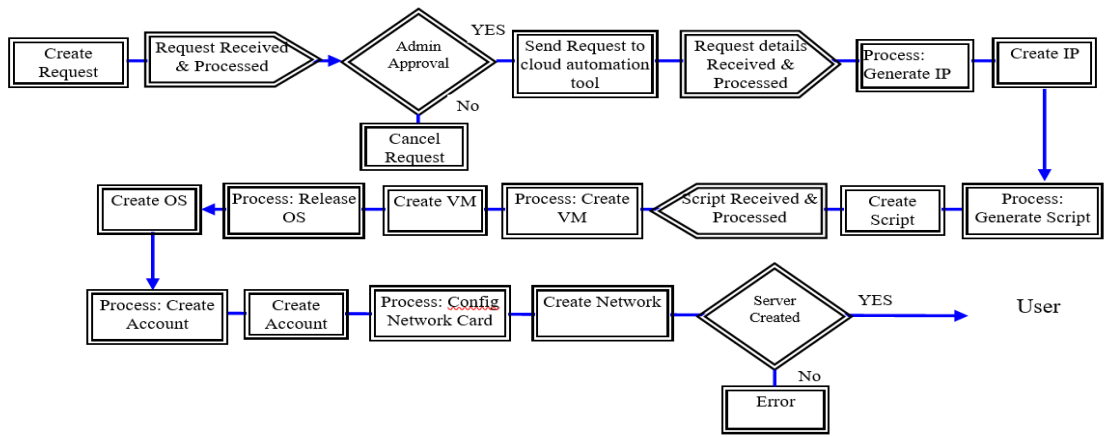

Fig. 14. Arena flowchart of the server request system discussed in IV.3

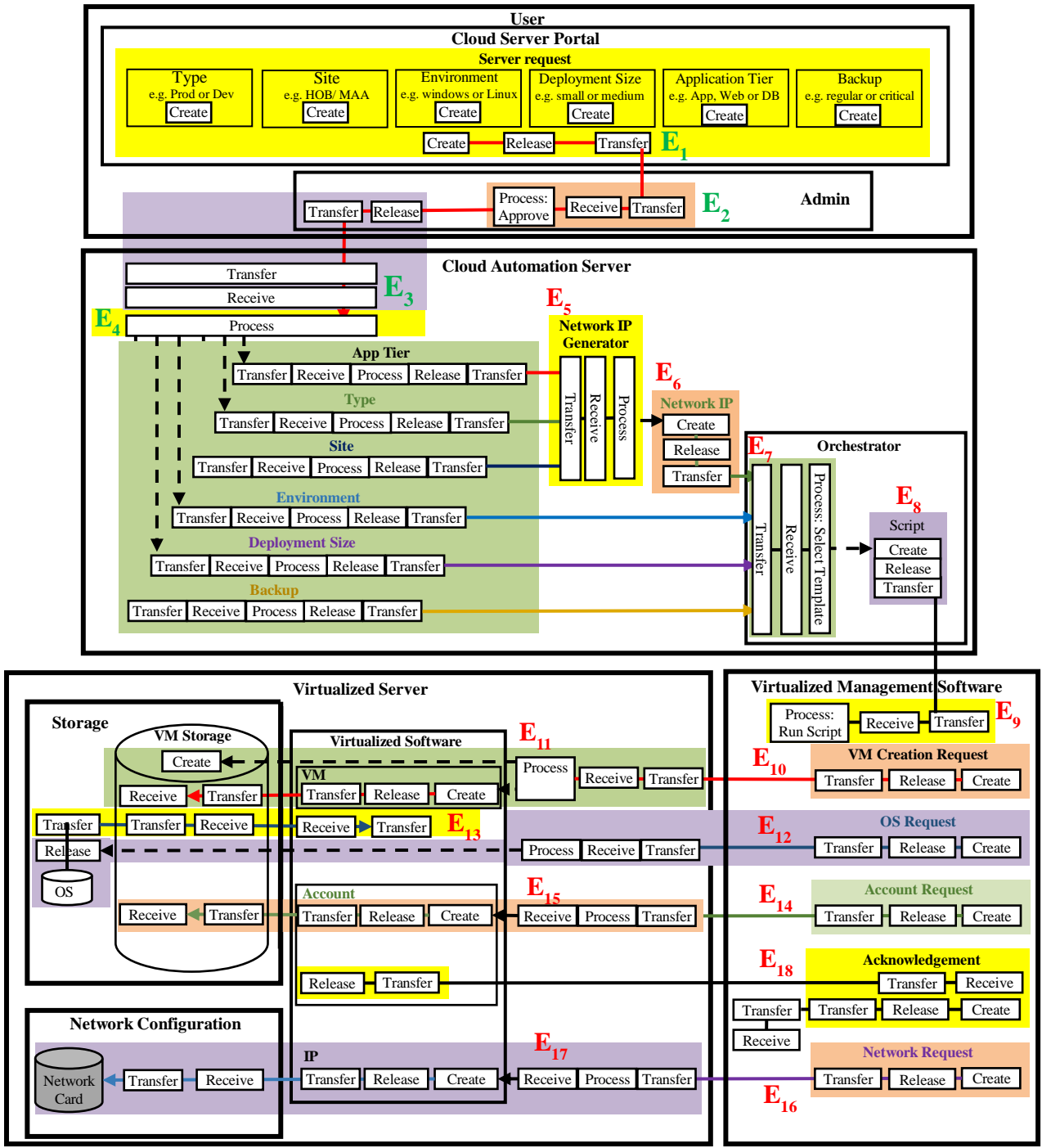

Fig. 16. Server request model

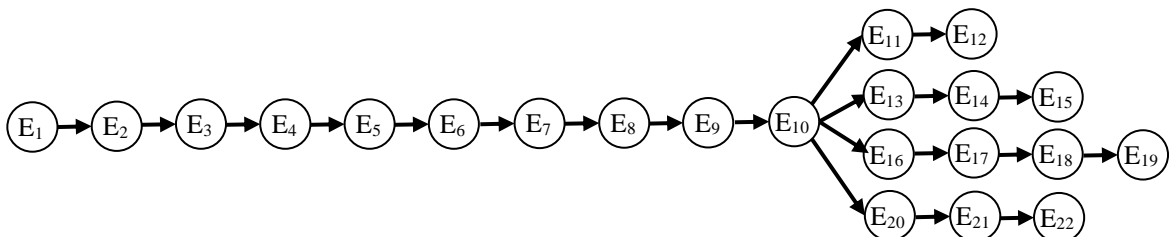

Fig. 17. Events

Event 4 (E4): The cloud automation tool processes the details received.

- Event 5 (E5): The app tier, type, and site are extracted to the network IP generator.
- Event 6 (E6): The network IP generator creates the IP.
- Event 7 (E7): The environment, deployment size, backup priority, and IP are released and processed by the cloud automation tool and sent to the orchestrator.
- Event 8 (E8): The orchestrator creates a script with the server specifications.
- Event 9 (E9): The script is sent to the virtual management software.
- Event 10 (E10): The script will then trigger the creation of a VM, OS, account, and network.
- Event 11 (E11): The VM creation request is transferred to the virtualized server.
- Event 12 (E12): The VM and the VM storage are created on the server.
- Event 13 (E13): The OS request is transferred to the virtualized software.
- Event 14 (E14): The OS is downloaded from the storage to the VM storage.
- Event 15 (E15): The OS is installed on the server.
- Event 16 (E16): The account request is transferred to the virtualized software.
- Event 17 (E17): The username creation is triggered, causing the creation of a username and password that will be stored in the VM storage.
- Event 18 (E18): The account creation triggers the creation of an acknowledgment.

- Event 19 (E19): The acknowledgement is transferred to the virtualized management software and then to the user.
- Event 20 (E20): The network request is transferred to the virtualized software.
- Event 21 (E21): The IP is created.
- Event 22 (E22): The IP is configured in the network card.

Mapping the Arena flowchart to the TM events as shown in Fig. 18 starts as follows:

- The first box, "Create Request," and the arrow in the flowchart correspond to Event 1 in the TM, as shown in Fig. 18.
- "Request, Receive, & Process," in addition to the arrow and comparison in the flowchart, correspond to Event 2 in the TM.
- "Send request to cloud automation & process" and the arrow in the flowchart correspond to Events 2 and 3.

These contrasts between the first three or four notions in the flowchart show how the different activities in the flowchart are mixed (e.g., Send Request to cloud automation & Process) and repeated (e.g., Admin approval in a process and the decision diamond). Some events are represented by the arrow (e.g., the arrow after Create). There is clearly no systematic thinking in drawing the flowchart in Fig. 14. Instead, the TM diagram can be taken as a foundation for identifying events in the simulated system based on its basic events. This will be studied in a sequel paper that deals with simulating networks.

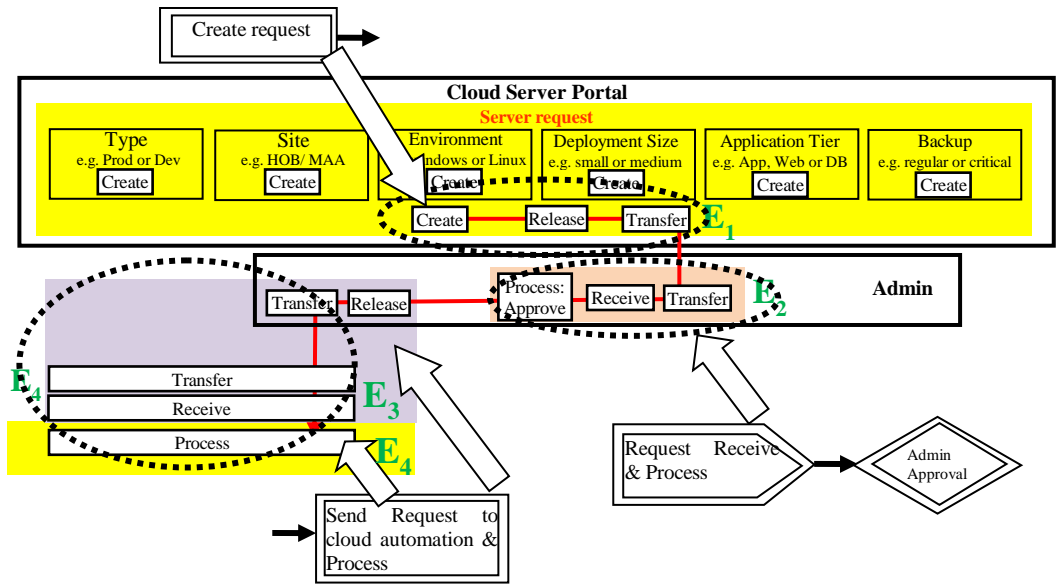

Fig. 18. Pointing out arbitrariness in the Arena flowchart using TM

## 5. Conclusion

This paper has demonstrated that a new model—the TM—can serve as a conceptual framework to give uniformity to computer networks. The TM diagram can be used at various levels of granularity and complexity, as in the case of nontechnical use. The viability of the model is demonstrated by applying it to actual cloud architecture.

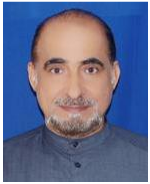

**Sabah S. Al-Fedaghi** is an associate professor in the Department of Computer Engineering at Kuwait University. He holds an MS and a PhD from the Department of Electrical Engineering and Computer Science, Northwestern University, Evanston, Illinois, and a BS in from Arizona State University. He has published more than 325 journal articles and papers in conferences on software engineering, database systems, information ethics, privacy, and security. He headed the Electrical and Computer Engineering Department (1991–1994) and the Computer Engineering Department (2000–2007). He previously worked as a programmer at the Kuwait Oil Company.

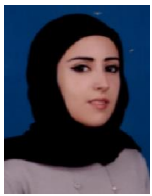

**Dana S. Al-Qemlas** received her BS degree in computer engineering from Kuwait University in 2013. She is currently pursuing an MS degree in Computer Engineering at Kuwait University. Dana's academic interests include computer architecture, virtualization, and computer networking